\documentclass[draftcls,onecolumn,11pt]{IEEEtran}

\usepackage{cite}

\ifCLASSINFOpdf
\else
  \usepackage[dvips]{graphicx}
  \graphicspath{{../}}
  \DeclareGraphicsExtensions{.eps}
\fi

\usepackage{amsmath,amsfonts}

\usepackage{algorithm}
\usepackage{algorithmic}

\ifCLASSOPTIONcompsoc
  \usepackage[caption=false,font=normalsize,labelfont=sf,textfont=sf]{subfig}
\else
  \usepackage[caption=false,font=footnotesize]{subfig}
\fi

\newtheorem{proposition}{Proposition}

\begin{document}

\title{Self-Organization Scheme for Balanced Routing in Large-Scale Multi-Hop Networks}


\author{Mihai-Alin~Badiu,
        David~Saad,
        and~Justin P. Coon,~\IEEEmembership{Senior~Member,~IEEE}
\thanks{M.-A Badiu is with the Department of Electronic Systems, Aalborg University, Aalborg 9220, Denmark (e-mail: mib@es.aau.dk) and the Department of Engineering Science, University of Oxford, Parks Road, Oxford OX1 3PJ, UK (email: mihai.badiu@eng.ox.ac.uk).}
\thanks{D. Saad is with the School of Engineering and Applied Science, Aston University, Birmingham B4 7ET, United Kingdom (e-mail: d.saad@aston.ac.uk).}
\thanks{J. P. Coon is with the Department of Engineering Science, University of Oxford, Parks Road, Oxford OX1 3PJ, UK (email: justin.coon@eng.ox.ac.uk).}
\thanks{This work was supported by the Independent Research Fund Denmark under Grant ID DFF--5054-00212, the Leverhulme Trust Grant No. RPG-2013-48 (D.S.) and EPSRC grant number EP/N002350/1 (``Spatially Embedded Networks''). Most of the research was carried out during an extended visit to Aston University, and M.-A Badiu would like to thank the Mathematics group for the hospitality.}
}

\maketitle

\begin{abstract}
We propose a self-organization scheme for cost-effective and load-balanced routing in multi-hop networks. To avoid overloading nodes that provide favourable routing conditions, we assign each node with a cost function that penalizes high loads. Thus, finding routes to sink nodes is formulated as an optimization problem in which the global objective function strikes a balance between route costs and node loads. We apply belief propagation (its min-sum version) to solve the network optimization problem and obtain a distributed algorithm whereby the nodes collectively discover globally optimal routes by performing low-complexity computations and exchanging messages with their neighbours. We prove that the proposed method converges to the global optimum after a finite number of local exchanges of messages. Finally, we demonstrate numerically our framework's efficacy in balancing the node loads and study the trade-off between load reduction and total cost minimization.
\end{abstract}


\section{Introduction}
\IEEEPARstart{L}{arge-scale} wireless networks employing multi-hop transmissions are an integral component of the Internet of Things~\cite{AlFuqaha2015}. For example, such networks can consist of a massive number of sensors that collect data from the environment and send it to central controllers. Since in multi-hop networks each wireless node can relay other nodes' messages, it is highly relevant to direct the information flows from the source nodes to the destinations efficiently in terms of, e.g., energy consumption or reliability. Sending the flows along the minimim-cost paths towards the destinations potentially leads to overloading those nodes that provide favourable routes, which can cause quick battery depletion or decrease the resilience of the network against node failures~\cite{Toh2001,Chang2004,Madan2006,Vazifehdan2014}. Therefore, information should be routed through the network so as to minimize costs while trying to balance the node-loads. Moreover, given their scale, such networks must be designed to be self-organizing and adaptive. 

There is a large body of work studying energy efficient routing protocols (see, e.g., the survey~\cite{Pantazis2013}). A typical objective is to maximize the network lifetime by maximizing the minimum lifetime over all nodes, where the lifetime of a node is defined as the ratio between its residual energy and its energy expenditure~\cite{Toh2001,Chang2004,Madan2006}. However, the network lifetime objective does not account for the total routing cost (total energy in this case) and thus can be inefficient in this respect, similar to minimum-cost routing being suboptimal for node balancing. It is therefore relevant to investigate objectives that favour solutions that are somewhere ``in-between'' these two extremes.

In this work, we propose an algorithmic strategy for distributed multi-hop networking whereby the nodes coordinate and organize themselves so as to route the information to the destinations in an efficient and balanced way. To this end, we model balanced routing as the minimization of a network objective function, which includes the overall cost of the routes (given by generic link costs) and an additional term that penalizes the node-loads. The objective function provides a tunable trade-off between total cost efficiency and fairness of the distribution of the node loads. The possible routes from source nodes to destinations are coupled in the objective function, which creates a competition for the shortest (i.e., least cost) routes to the sinks. To solve the optimization problem, we use the min-sum version of the belief propagation (BP) method~\cite{Mezard2009}. In this way, we obtain a distributed algorithm which finds globally optimal routes in a decentralized manner with low-complexity local computations and message exchanges between neighbouring nodes. We also show that the proposed method converges to the global optimum in a finite number of iterations.

\section{Network Model and Problem Formulation}

We assume a data collection scenario in which a set $\mathcal{V}_{\text{s}}=\{1,\ldots,n\}$ of $n$ nodes generate and/or relay information that has to be delivered to any subset of the $m$ destination nodes (e.g., gateways, access points) in $\mathcal{V}_{\text{d}}=\{n+1,\ldots,n+m\}$. The nodes in $\mathcal{V}_{\text{s}}$ are simple devices with constrained resources (energy, memory, processing capabilities, etc.) and can participate in routing each other's packets towards the destination nodes. Packets generated by a source node in $\mathcal{V}_{\text{s}}$ can travel to a destination in $\mathcal{V}_{\text{d}}$ over different routes; moreover, they can be delivered to different destination nodes. 

We model the wireless network as a directed graph $G(\mathcal{V},\mathcal{E})$, with $\mathcal{V}=\mathcal{V}_{\text{s}}\cup \mathcal{V}_{\text{d}}$ and $\mathcal{E}$ being the set of edges (links). An edge $(i,j)\in\mathcal{E}$ indicates that node $i$ can transmit to node $j$ directly. For each $i\in\mathcal{V}$, $\mathcal{E}_i$ denotes the set of all edges incident to $i$, while $\mathcal{E}_i^{\text{out}}$ and $\mathcal{E}_i^{\text{in}}$ stand for the sets of its outgoing and respectively incoming edges. Node $i\in\mathcal{V}_{\text{s}}$ generates information at a rate of $r_i$ units (we assume a certain unit rate $[r]$), where $r_i\in\mathbb{N}$; if $r_i=0$, the node is just a relay node. The capacity of edge $e\in\mathcal{E}$ is $u_e$ units, $u_e\in\mathbb{N}_{>0}$, such that the amount of flow $x_e$ units carried by $e$ satisfies $0\leq x_e\leq u_e$. The assumption that the rates and capacities are integer multiples of $[r]$ is not restrictive, because any set of rational numbers can be expressed in this way by finding an appropriate unit $[r]$. Moreover, if any of the rates and capacities have irrational values, it is necessary to convert them to rational numbers to represent them on a computer. We associate each link $e\in\mathcal{E}$ with the weight $c_e>0$ representing the cost of transferring a unit over edge $e$. For example, the cost can be the transmit power required to ensure a certain data rate, the expected transmission count (ETX), or hop-count (when $c_e=1$). We further assume that the network is in the unsaturated traffic regime and packets are transferred between neighbours according to a medium access scheme, which we do not concern ourselves with here.

The routing solution space consists of those configurations $\{x_e\}_{e\in\mathcal{E}}$ which satisfy the flow conservation constraints
\begin{equation}\label{eq:flow_cons}
\sum_{e\in\mathcal{E}_i^{\text{out}}} x_e - \sum_{e\in\mathcal{E}_i^{\text{in}}} x_e = r_i,\quad \text{for all } i\in\mathcal{V}_{\text{s}},
\end{equation}
and the capacity constraints $0\leq x_e \leq u_e$, for all $e\in\mathcal{E}$. The two constraints ensure that all generated flows are delivered to the destinations such that edge flows do not exceed the respective capacities. We assume that the solution space is non-empty. The total cost of a configuration $\{x_e\}_{e\in\mathcal{E}}$ is $\sum_{e\in\mathcal{E}} c_e x_e$. Furthermore, we define the load of node $i$ to be the amount of flow $\sum_{e\in\mathcal{E}_i^{\text{out}}} x_e$ it has to forward.

In general, there are many feasible configurations, each implying different sets of routes, path lengths, total costs, distribution of node loads, etc.
A common objective is to minimize the total cost, which, as one can notice, turns data collection into a (linear) minimum cost network flow problem~\cite{Ahuja1993}. However, such an approach may yield solutions wherein some nodes that provide low-cost forwarding edges experience high loads. We are therefore interested in balancing the node loads in a cost-effective manner.

\section{Proposed Objective for Load Balancing}
We seek a trade-off between minimization of the total cost and minimization of the loads of individual nodes. To this end, for each $i\in\mathcal{V}_{\text{s}}$ we introduce the strictly-increasing convex function $\phi_i:[0,\infty)\to\mathbb{R}$ to penalize the load of the $i$th node. The functions can vary over the nodes to reflect their different load-tolerances depending on residual energies, capabilities etc. Now, we formulate the optimization problem
\begin{equation}\label{eq:obj_balance}
\begin{aligned}
    &\underset{\mathbf{x}\in\mathbb{R}^{|\mathcal{E}|}}{\text{minimize}} && (1-w) \sum_{e\in\mathcal{E}} c_e x_e + w\sum_{i\in\mathcal{V}_{\text{s}}} \phi_i\left(\sum\nolimits_{e\in\mathcal{E}_i^{\text{out}}} x_e\right) \\
    &\text{subject to} && \sum_{e\in\mathcal{E}_i^{\text{out}}} x_e - \sum_{e\in\mathcal{E}_i^{\text{in}}} x_e = r_i,\quad \forall i\in\mathcal{V}_{\text{s}}, \\
    &&& 0\leq x_e \leq u_e,\quad \forall e\in\mathcal{E},
\end{aligned}
\end{equation}
where $w$ is a parameter that balances cost-efficiency and load minimization. When $w=0$, we recover the linear minimum cost flow problem~\cite{Ahuja1993}, which gives the most cost-efficient flow configuration; however, this setting usually does not provide well-balanced loads and therefore we focus on $w>0$.

In the following, we assume that the functions $\phi_i$ are piecewise-linear convex (PLC) with integral breakpoints, which is very convenient for obtaining a simple message-passing algorithm with provable convergence to the correct solution, as we show next in Prop.~\ref{prop:Complexity} and Prop.~\ref{prop:Convergence}. An example of such function is one that takes the value $y^\alpha$, with $\alpha>1$, at each breakpoint $y\in\mathbb{N}$ and varies linearly between consecutive breakpoints; the higher the value of $\alpha$, the stronger the load $y$ is penalized. Such a choice provides a simple way to select the efficiency-fairness trade-off by tuning the parameter $\alpha$.

\section{BP Algorithm for Balanced Routing}
BP is a generic message-passing algorithm for solving large-scale inference and optimization problems in graphical models. It has a distributed nature whereby the nodes of the graph perform simple local computations and exchange messages with their neighbours. While BP provides correct solutions when the underlying graph is a tree, its correctness and convergence cannot be generally guaranteed for graphs with cycles, with few exceptions~\cite{Mezard2009,Gamarnik2012}. Nonetheless, for graphs with cycles, the BP heuristic often performs very well. 
In network problems, the min-sum algorithm is applied to find the shortest path between two nodes~\cite{Ruozzi2008} or minimize path lengths and link congestion~\cite{Yeung2012}. For the min-cost network flow problem with linear or PLC costs on edges, BP was shown in~\cite{Gamarnik2012} to converge to the correct solution (if the solution is unique). Compared to~\cite{Gamarnik2012}, our objective~\eqref{eq:obj_balance} (with $w>0$) additionally includes node costs given by the PLC functions $\{\phi_i\}$; therefore, the application of BP gives the novel algorithm described next.\footnote{\label{footnote:BPsplit}Alternatively, by using the node splitting technique~\cite[p.~41]{Ahuja1993}, one can transform~\eqref{eq:obj_balance} into a min-cost network flow problem with PLC costs on edges, which can be solved using BP~\cite[Th. 6.1]{Gamarnik2012}. However, BP on the transformed graph is different from Algorithm~\ref{alg:BP_routing} that we obtain here, see footnote~\ref{footnote:msgBPsplit}.}

For each node $i\in\mathcal{V}_{\text{s}}$, we define a function $\psi_i$ that reflects the flow conservation constraint at node $i$, i.e., it maps each vector $\mathbf{z}\in\mathbb{R}_{+}^{|\mathcal{E}_i|}$ of edge flows to
\begin{equation*}
    \psi_i(\mathbf{z}) =
    \begin{cases}
    0, & \text{if} \quad \sum\limits_{e\in\mathcal{E}_i^{\text{out}}} z_e - \sum\limits_{e\in\mathcal{E}_i^{\text{in}}} z_e = r_i, \\
    \infty, & \text{otherwise}.
    \end{cases}
\end{equation*}
Furthermore, we define
\begin{equation*}
    f_i(\mathbf{z}) = \psi_i(\mathbf{z}) + w\,\phi_i\left(\sum\nolimits_{e\in\mathcal{E}_i^{\text{out}}} z_e \right),
\end{equation*}
which additionally includes the load penalty for node $i\in\mathcal{V}_{\text{s}}$. On the contrary, destination nodes do not have any constraints and ``accept'' any flows on their incoming edges, so we set $f_i(\mathbf{z})=0$, for any $i\in\mathcal{V}_{\text{d}}$ and $\mathbf{z}\in\mathbb{R}_{+}^{|\mathcal{E}_i|}$. Next, we capture the cost and capacity constraint of edge $e\in\mathcal{E}$ by introducing the function $g_e:\mathbb{R}\to\mathbb{R}\cup\{\infty\}$ given by
\begin{equation*}
    g_e(z) \\ =
    \begin{cases}
    (1-w)c_e z, & \text{if} \quad 0\leq z \leq u_e, \\
    \infty, & \text{otherwise}.
    \end{cases}
\end{equation*}
We can now reformulate~\eqref{eq:obj_balance} as the equivalent problem
\begin{equation}\label{eq:obj_BP}
\underset{\mathbf{x}\in\mathbb{R}^{|\mathcal{E}|}}{\text{minimize}} \sum_{e\in\mathcal{E}} g_e(x_e) + \sum_{i\in\mathcal{V}} f_i(\mathbf{x}_{\mathcal{E}_i}),
\end{equation}
where $\mathbf{x}_{\mathcal{E}_i}$ includes those components of $\mathbf{x}$ with indices in $\mathcal{E}_i$.

We apply the min-sum version of BP to solve~\eqref{eq:obj_BP}. Given that each edge variable node has exactly two neighbour function nodes from the set $\{f_i\}$, we simplify the standard message updates by defining the messages~\eqref{eq:msg_update} in  Algorithm~\ref{alg:BP_routing}. At iteration $t$, for each node $i\in\mathcal{V}_{\text{s}}$ and incident edge $e\in\mathcal{E}_i$, where either $e=(i,j)\in\mathcal{E}_i^{\text{out}}$ or $e=(j,i)\in\mathcal{E}_i^{\text{in}}$, the algorithm computes the message $m_{i\to e}^t$, which becomes an input to neighbour $j$ at the next iteration. Since $f_i$ is the zero function for all $i\in\mathcal{V}_{\text{d}}$, the messages computed by destination nodes do not change with $t$ and thus are not updated.

\begin{algorithm}[H]
\caption{Distributed algorithm for balanced routing.}
\label{alg:BP_routing}
\begin{algorithmic}[1]
\renewcommand{\algorithmicrequire}{\textbf{Input:}}
\renewcommand{\algorithmicensure}{\textbf{Output:}}
\REQUIRE The graph $G(\mathcal{V},\mathcal{E})$, edge costs $\{c_e\}$ and capacities $\{u_e\}$, data rates $\{r_i\}$, parameters $\alpha$ and $w$
\ENSURE Estimates $\{\hat{x}_e\}_{e\in\mathcal{E}}$ of the optimal edge flows of~\eqref{eq:obj_balance}
\STATE{Initialize $m_{i\to e}^0(z) = g_e(z)$, for all $i\in\mathcal{V}$, $e\in\mathcal{E}_i$, $z\in\mathbb{R}_{+}$ }
\FOR{$t=1$ \TO $T$}
    \STATE{For each $i\in\mathcal{V}_{\text{s}}$ and $e\in\mathcal{E}_i$, update
    \begin{equation}\label{eq:msg_update}
    m_{i\to e}^t(z) = g_e(z) + \min_{\mathbf{\tilde{z}}\in\mathbb{R}_+^{|\mathcal{E}_i|}: \tilde{z}_e=z} \left\{f_i(\mathbf{\tilde{z}}) + \sum_{e'\in\mathcal{E}_i\setminus e}  m_{k\to e'}^{t-1}(\tilde{z}_{e'})\right\},
    \end{equation}
    for all $z\in\mathbb{R}_{+}$}, where $e'=(i,k)\text{ or }(k,i)$.
\ENDFOR
\STATE{For each $e=(i,j)\in\mathcal{E}$, compute the belief function
\begin{equation}\label{eq:belief}
b_e^t(z) = m_{i\to e}^t(z) + m_{j\to e}^t(z) - g_e(z)
\end{equation}
and determine its minimizer
\begin{equation}\label{eq:min_belief}
\hat{x}_e^t = \operatorname*{arg\,min}_z  b_e^t(z)
\end{equation}
}
\RETURN{$\mathbf{\hat{x}}^t=\left(\hat{x}_e^t\right)_{e\in\mathcal{E}}$}
\end{algorithmic}
\end{algorithm}

\begin{figure*}[!t]
\centering
\subfloat[]{
    \includegraphics[width=0.31\textwidth]{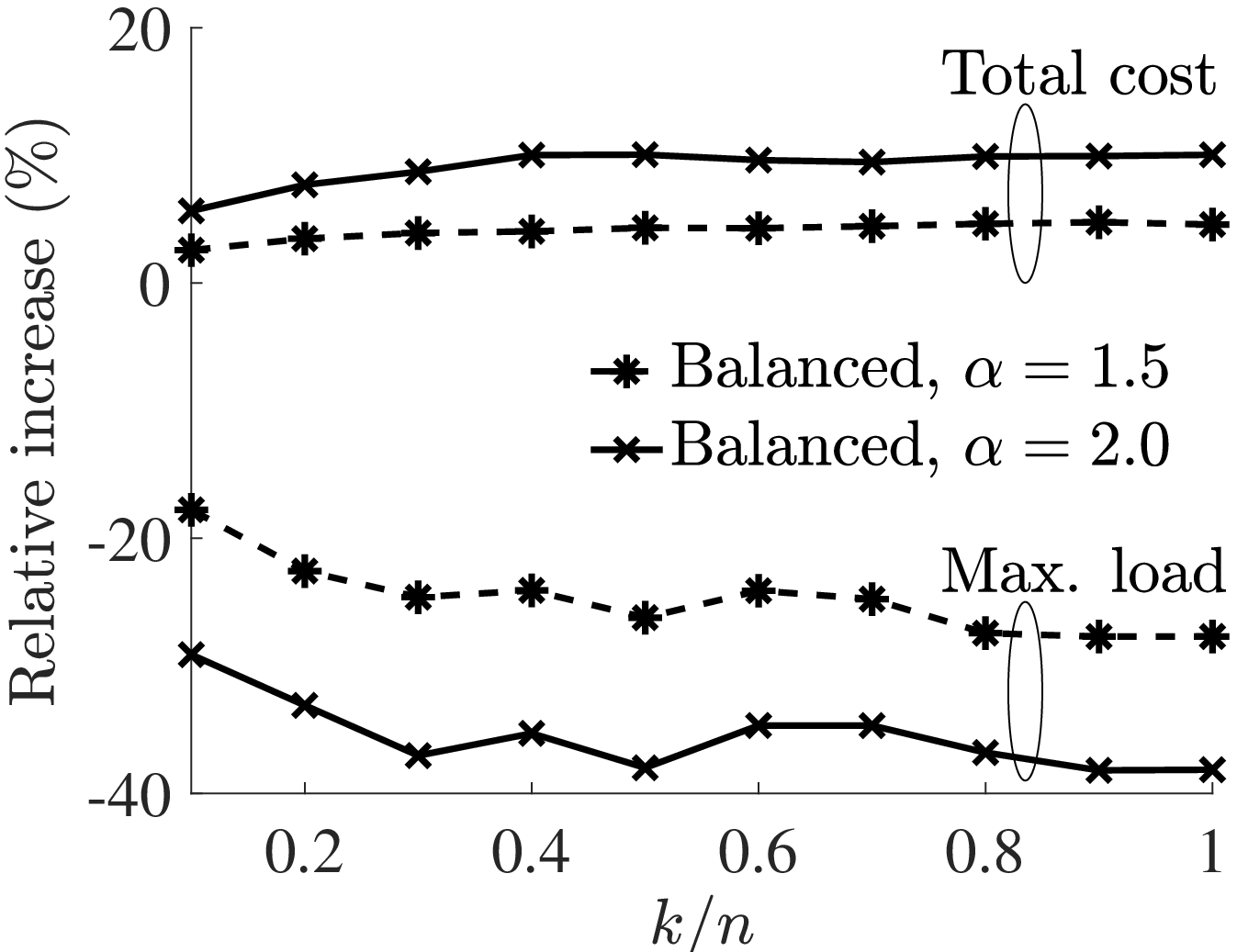}%
    \label{fig:RelativeImprovement}
}
\hfil
\subfloat[]{
    \includegraphics[width=0.31\textwidth]{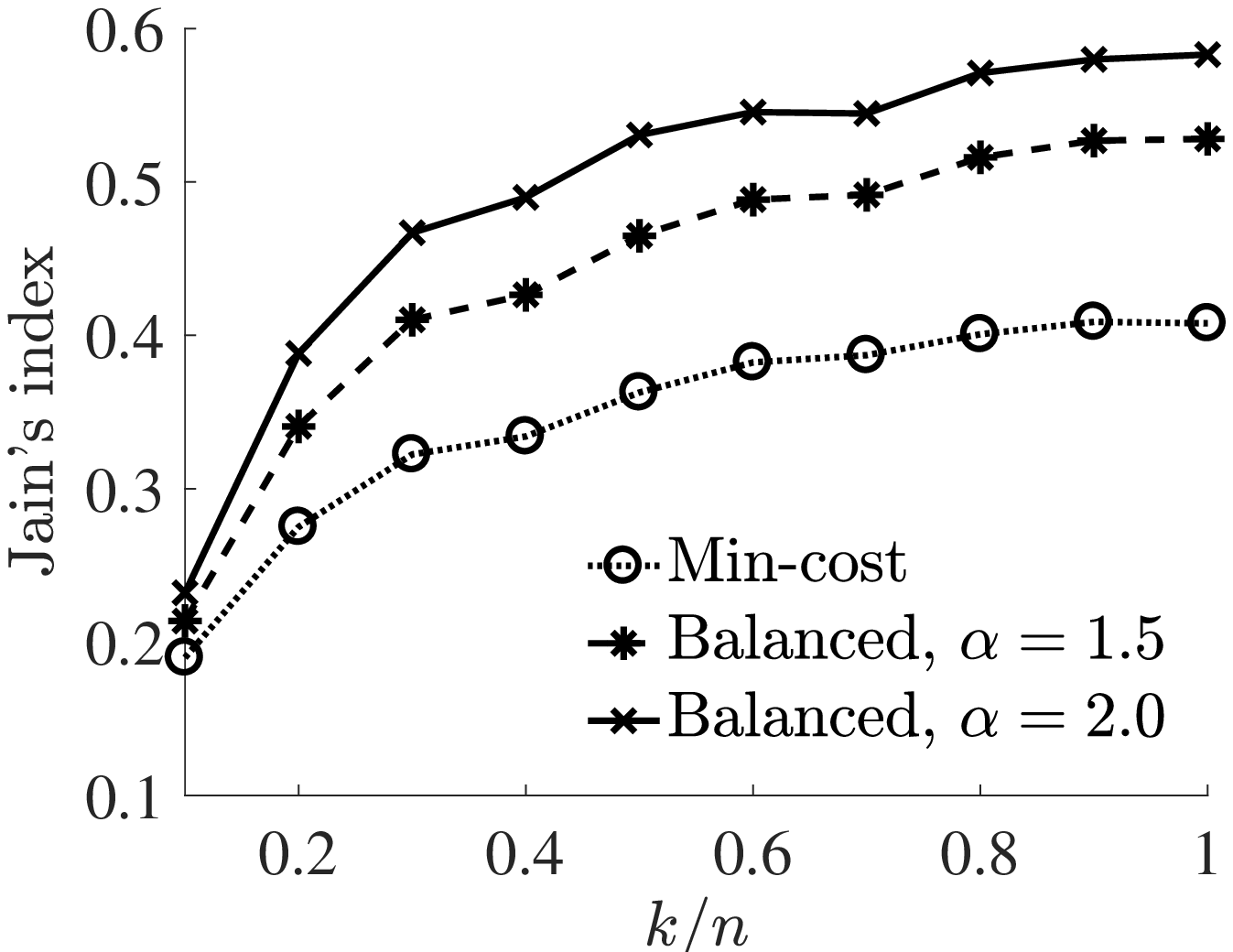}%
    \label{fig:Jain_index}
}
\hfil
\subfloat[]{
    \includegraphics[width=0.31\textwidth]{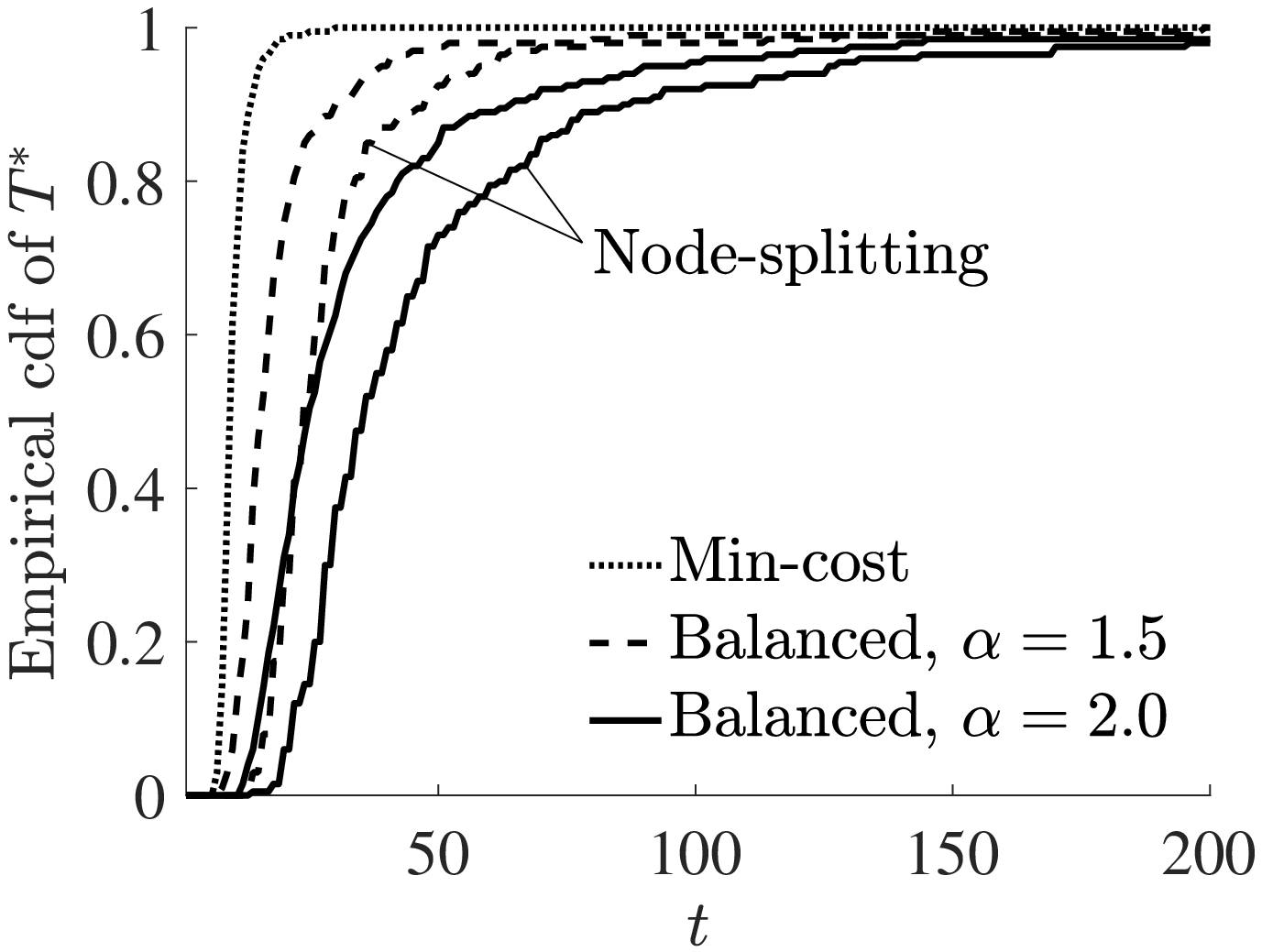}%
    \label{fig:Iterations}
}
\caption{Simulation results for $n=50$, $m=1$ and various fractions $k/n$ of source nodes: (a) Improvement of the total cost and maximum load relative to minimum-cost routing; (b) Jain's fairness index for the node loads; (c) empirical cdf of the minimum number $T^\ast$ of iterations required for Algorithm~\ref{alg:BP_routing} to converge when $k/n=0.3$.}
\label{fig:Results}
\end{figure*}

Algorithm~\ref{alg:BP_routing} has the following interpretation. Every node is seeking to determine the flow on each of its incident edges while satisfying its local flow conservation constraint and minimizing its load. The message $m_{i\to e}^t(z)$ can be viewed as a local cost that node $i$ attributes to allocating $z$ units to edge $e$; thus, the message is a function of the flow. For any $z$, the message update~\eqref{eq:msg_update} includes: (i) the cost of sending flow $z$ over edge $e$ and (ii) the minimum cost of allocating flows to the rest of the edges that are incident to $i$ such that flow conservation is ensured. The latter cost is the result of a local optimization, which looks for the feasible configuration of the flows on the incident edges that minimizes an objective function that includes the cost of the load of node $i$ and the local costs (messages) estimated by the neighbouring nodes.\footnote{\label{footnote:msgBPsplit}If using BP on the augmented graph obtained by node-splitting (see footnote~\ref{footnote:BPsplit}), then $m_{i\to e}^t$ effectively depends on messages over incoming (if $e\in\mathcal{E}_i^{\text{out}}$) or outgoing (if $e\in\mathcal{E}_i^{\text{in}}$) edges that were computed at $t-2$, i.e., it uses outdated information, which slows down convergence, as Fig.~\ref{fig:Iterations} shows.} The message updates have low-complexity, as we show next.
\begin{proposition}[Complexity]\label{prop:Complexity}
For each $i\in\mathcal{V}_{\text{s}}$, $e\in\mathcal{E}_i$ and $t\geq 1$, the message $m_{i\to e}^t$ is a piecewise-linear convex (PLC) function with breakpoints in $\{0,1,\ldots,u_e\}$. The complexity of its update~\eqref{eq:msg_update} is linear in the total capacity of the input and output edges of node $i$ and logarithmic in $|\mathcal{E}_i^{\text{in}}|$ and $|\mathcal{E}_i^{\text{out}}|$.
\end{proposition}
\begin{IEEEproof}
The proof is by induction on $t$. At $t=0$, Algorithm~\ref{alg:BP_routing} initializes the messages to trivial PLC functions. Suppose at iteration $t-1$ all messages are PLC functions with integral breakpoints.
We provide the proof for $m_{i\to e}^t$ with $e\in\mathcal{E}_i^{\text{out}}$, as the case $e\in\mathcal{E}_i^{\text{in}}$ is very similar. Let $\psi_y^{(1)}:\mathbb{R}^{\left|\mathcal{E}_i^{\text{in}}\right|} \to \mathbb{R}\cup\{\infty\}$, $y\in\mathbb{R}$, be
\begin{equation*}
    \psi_y^{(1)}(\mathbf{z}) =
    \begin{cases}
    0, & \text{if} \quad \sum\limits_{e'\in\mathcal{E}_i^{\text{in}}} z_{e'} + r_i = y, \\
    \infty, & \text{otherwise},
    \end{cases}
\end{equation*}
and define $\psi_v^{(2)}:\mathbb{R}^{\left|\mathcal{E}_i^{\text{out}}\right|-1}\times\mathbb{R} \to \mathbb{R}\cup\{\infty\}$, $v\in\mathbb{R}$, given by
\begin{equation*}
    \psi_v^{(2)}(\mathbf{z},y) =
    \begin{cases}
    0, & \text{if} \quad y - \sum\limits_{e'\in\mathcal{E}_i^{\text{out}}\setminus e} z_{e'} = v, \\
    \infty, & \text{otherwise}.
    \end{cases}
\end{equation*}
Now, we define the function
\begin{equation*}
  h(y) = \min_{\mathbf{\tilde{z}}\in \mathbb{R}^{\left|\mathcal{E}_i^{\text{in}}\right|}} \left\{ \psi_y^{(1)}(\mathbf{\tilde{z}}) + \sum_{e'=(k,i)\in\mathcal{E}_i^{\text{in}}}  m_{k\to e'}^{t-1}(\tilde{z}_{e'}) \right\}+ w\phi_i(y).
\end{equation*}
The minimization in the r.h.s. is a so-called interpolation of PLC functions whose complexity is logarithmic in the number of functions and linear in the total number of their linear pieces~\cite{Gamarnik2012}. Since $m_{k\to e'}^{t-1}$ has breakpoints in $\{0,1,\ldots,u_{e'}\}$ and $\phi_i$ is also PLC with integral breakpoints, it follows that the function $h$ is itself PLC with integral breakpoints and at most $U_i^{\text{in}}$ pieces, where $U_i^{\text{in}} = \sum_{e'\in\mathcal{E}_i^{\text{in}}} u_{e'}+r_i$; moreover, $h$ can be computed in $O(U_i^{\text{in}}\log|\mathcal{E}_i^{\text{in}}|)$ operations. Now, we write~\eqref{eq:msg_update}
\begin{equation*}
  m_{i\to e}^t(z) = g_e(z) + \min_{\mathbf{\tilde{z}},y} \left\{ \psi_z^{(2)}(\mathbf{\tilde{z}},y)+ h(y) + \sum_{e'=(i,k)\in\mathcal{E}_i^{\text{out}}\setminus e}  m_{k\to e'}^{t-1}(\tilde{z}_{e'}) \right\}.
\end{equation*}
Given that $h$ and the messages at $t-1$ are PLC with integral breakpoints, the interpolation in the second line gives again a PLC function; its computation takes $O\left(U_i^e\log|\mathcal{E}_i^{\text{out}}|\right)$ operations, where $U_i^e = \sum_{e'\in\mathcal{E}_i\setminus e} u_{e'}$. The addition of $g_e$, which is linear in $[0,u_e]$, makes $m_{i\to e}^t$ PLC with integral breakpoints.
\end{IEEEproof}
We establish that Algorithm~\ref{alg:BP_routing} outputs the optimal solution after a finite number of iterations.
\begin{proposition}[Convergence]\label{prop:Convergence}
Suppose~\eqref{eq:obj_balance} has a unique optimal solution $\mathbf{x}^\ast$.\footnote{When the costs $\{c_e\}$ are generic (e.g., random), it is highly likely that~\eqref{eq:obj_balance} has a unique solution. Otherwise, it is possible to add small noise to the costs such that the modified problem has a unique solution which very closely approximates the solution of the original problem~\cite{Gamarnik2012}.} Then, there exists a finite integer $T^\ast$ such that the output of Algorithm~\ref{alg:BP_routing} satisfies $\mathbf{\hat{x}}^t=\mathbf{x}^\ast$, for any $t\geq T^\ast$.
\end{proposition}
\begin{IEEEproof}
Although our objective function~\eqref{eq:obj_balance} is different than that of the min-cost network flow problem with linear (or PLC) edge costs, we can use the same proof strategy as in~\cite[Th.~4.1, Th.~6.1]{Gamarnik2012}. The difference is that we need to define an appropriate residual graph~\cite{Ahuja1993}. Denote by $G(\mathbf{x})$ the residual graph of $G(\mathcal{V},\mathcal{E})$ with respect to the flow $\mathbf{x}\in\mathbb{R}^{|\mathcal{E}|}$. $G(\mathbf{x})$ has the same vertices $\mathcal{V}$, while we define its edges and their costs as follows: for any $e=(i,j)\in\mathcal{E}$, if $x_e<u_e$, then $e$ is also an edge in $G(\mathbf{x})$ with capacity $u_e-x_e$ and cost $c_e^{\mathbf{x}}=(1-w)c_e+w\lim_{z\rightarrow 0^{+}} \left(\phi_i(y+z)-\phi_i(y)\right)/z$, where $y=\sum_{e'\in\mathcal{E}_i^{\text{out}}} x_{e'}$ is the load of node $i$; if $x_e>0$, then $G(\mathbf{x})$ additionally includes the directed edge $e'=(j,i)$ with capacity $x_e$ and cost $c_{e'}^{\mathbf{x}}=-(1-w)c_e+w\lim_{z\rightarrow 0^{-}} \left(\phi_i(y+z)-\phi_i(y)\right)/z$. At the unique optimal solution $\mathbf{x}^\ast$, all the directed cycles of the residual graph $G(\mathbf{x}^\ast)$ must have positive costs (according to the negative cycle optimality criterion~\cite{Ahuja1993}). The proof relies on this property and follows the same steps as that of~\cite[Th. 4.1]{Gamarnik2012}; therefore we omit the details.
\end{IEEEproof}

\section{Numerical Results}

We consider $n=50$ nodes independently and uniformly distributed inside the unit square and $m=1$ sink node at the center of the square. Any two nodes $i$ and $j$ that are spaced by less than $1.6/\sqrt{n}\approx 0.23$ are connected by the directed edges $(i,j)$ and $(j,i)$. We discard the network realizations that are not connected. For each realization, we randomly select $k$ sources out of the $n$ nodes; the sources generate information at unit rate, while the remaining $n-k$ nodes act as relays. The cost associated with each link is the expected transmission count (ETX), which is drawn uniformly at random from the interval $[1,3]$. 

For the proposed balanced routing scheme (Algorithm~\ref{alg:BP_routing} with $w=0.5$ and power $\alpha>0$), we evaluate the total cost, the maximum of the node loads $\{y_i\}_{i=1}^n$, the  Jain's index, $J=\frac{1}{n}\left(\sum_{i=1}^n y_i\right)^2/\sum_{i=1}^n y_i^2 \in [1/n,1]$, as a measure of fairness in the distribution of the loads and the empirical distribution of the minimum number $T^\ast$ of iterations required for Algorithm~\ref{alg:BP_routing} to converge. We compare the results obtained using our algorithm against minimum-cost routing which is instantiated by setting $w=0$ in Algorithm~\ref{alg:BP_routing}. The results in Fig.~\ref{fig:Results} are obtained by averaging from $200$ independent trials. In Fig.~\ref{fig:RelativeImprovement}, we observe that balancing with $\alpha=1.5$ reduces the maximum load by $20$--$25\%$ compared to minimum-cost routing across all fractions of source nodes, while the total cost increases by $<5\%$; increasing $\alpha$ to two brings larger reduction of the maximum load, of about $30$--$40\%$, and a higher relative total cost of about $5$--$10\%$. Fig.~\ref{fig:Jain_index} shows that the balanced routing scheme provides significantly fairer load-distributions. As illustrated in Fig.~\ref{fig:Iterations}, the number of iterations required to find a balanced solution is higher than for min-cost routing and increases with $\alpha$. We also evaluated BP on the graph transformed by node splitting (see footnotes~\ref{footnote:BPsplit} and~\ref{footnote:msgBPsplit}) and, while it outputs the same solutions, it requires a higher number of iterations than our method, as shown in Fig.~\ref{fig:Iterations}. 

\section{Conclusion}
We formulated balanced routing in large-scale networks (such as Internet of Things) as optimization of an objective function that provides a tunable trade-off between total cost efficiency and fairness of the distribution of the node loads. In the proposed decentralized scheme, the nodes collectively find the globally optimal routing solution through low-complexity local computations and exchanges of messages with neighbours. The scheme provides significantly fairer solutions than minimum-cost routing at the expense of slightly increased total cost and higher number of required iterations. 

There are several interesting directions to explore further, such as adapting the framework to specific models of energy consumption, including in the design the notions of reliability, trust among nodes and security, but also extending the framework to take into account the scheduling of the transmissions.

\ifCLASSOPTIONcaptionsoff
  \newpage
\fi



\bibliographystyle{IEEEtran}
\bibliography{IEEEabrv,refs}
\end{document}